\documentclass[24pt, usenatbib]{mn2e}
\usepackage{graphicx,times}
\usepackage{bm}
\usepackage{epsf}
\usepackage{amsmath}
\usepackage{amssymb}
\usepackage{colortbl}
\usepackage{hyperref}
\usepackage{url}
\usepackage{subfig}

\def\cb{\textcolor{black}}
\def\la{\langle}
\def\ra{\rangle}

\def\be{\begin{equation}}
\def\ee{\end{equation}}
\def\ben{\begin{eqnarray}}
\def\een{\end{eqnarray}}

\def\oh{{\bf{\widehat\Omega}}}

\def\bk{{\bf k}}

\def\br{{\bf r}}

\def\bk{{\bf k}}

\def\bl{{\bf l}}

\def\2p{{(2\pi)^2}}

\def\bl{{\bf l}}
\def\be{\begin{equation}}
\def\ee{\end{equation}}
\def\beq{\begin{equation}}
\def\eeq{\end{equation}}
\def\ben{\begin{eqnarray}}
\def\een{\end{eqnarray}}
\def\bes{\begin{subequations}}
\def\ees{\end{subequations}}

\newcommand{\beqa}{\begin{eqnarray}}
\newcommand{\eeqa}{\end{eqnarray}}

\def\ikap0{{\cal J}_{\theta_0}(r)}

\def\one1{\langle \kappa_{(i)}\kappa_{(j)} \rangle}
\def\one{{[\bar \xi^{(ij)}]}}

\def\ba{\begin{eqnarray}}
\def\ea{\end{eqnarray}}

\def\bk{{\bf k}}

\def\br{{\bf r}}

\def\bk{{\bf k}}

\def\bl{{\bf l}}

\def\2p{{(2\pi)^2}}

\def\bl{{\bf l}}

\def\be{\begin{equation}}
\def\ee{\end{equation}}

\def\beq{\begin{equation}}
\def\eeq{\end{equation}}

\def\ben{\begin{eqnarray}}
\def\een{\end{eqnarray}}

\def\bk{{\bf k}}

\def\bl{{\mathbf l}}

\def\2p{{(2\pi)^2}}

\def\bl{{\bf l}}

\def\bl{{\bf{l}}}

\begin{document}
\onecolumn
\title[The Weak Lensing Bispectrum Induced By Gravity]
{The Weak Lensing Bispectrum Induced By Gravity}
\author[Munshi et al.]
       {D. Munshi$^{a}$, T. Namikawa$^b$, T. D. Kitching$^a$, J. D. McEwen$^a$, 
        \newauthor  
         R. Takahashi$^c$, F. R. Bouchet$^d$, A. Taruya$^{e}$,  B. Bose$^{f}$\\
        $^{a}$ Mullard Space Science Laboratory, University College London,
        Holmbury St Mary, Dorking, Surrey RH5 6NT, UK \\
        %
        %
        $^{b}$ Department of Applied Mathematics and Theoretical Physics, 
        University of Cambridge, Wilberforce Road, Cambridge CB3 OWA, UK
        \\
        $^{e}$ Faculty of Science and Technology, Hirosaki University,
        3 Bunkyo-cho, Hirosaki, Aomori, 036-8561, Japan\\
        %
        %
        $^{d}$ Institut d'Astrophysique de Paris, UMR 7095,
        CNRS \& Sorbonne Université, 98 bis Boulevard Arago, F-75014 Paris, France\\
        $^{e}$ Center for Gravitational Physics, Yukawa Institute for Theoretical Physics,
        Kyoto University, Kyoto 606-8502, Japan\\
        $^{f}$ Departement de Physique,  Universite de Geneve,
        24 quai Ernest-Ansermet, CH-1211 Geneve 4, Switzerland
       }
\maketitle
\begin{abstract}
  Recent studies have demonstrated that {\em secondary} non-Gaussianity induced by gravity will be detected with a high
  signal-to-noise (S/N) by future and even by on-going weak lensing surveys. One way to characterise such non-Gaussianity
  is through the detection of a non-zero three-point correlation function of the lensing convergence field, or of its harmonic transform,
  the bispectrum. A recent study analysed the properties of the squeezed configuration of the bispectrum, when two wavenumbers
  are much larger than the third one. We extend this work by estimating the amplitude of the (reduced) bispectrum in four generic configurations,
  i.e., {\em squeezed, equilateral, isosceles} and {\em folded}, and for four different source redshifts $z_s=0.5,1.0,1.5,2.0$,
  by using an ensemble of all-sky high-resolution simulations. We compare these results against theoretical predictions.
  We find that, while the theoretical expectations based on widely used fitting functions can predict the general trends of
  the reduced bispectra, a more accurate theoretical modelling will be required to analyse the next generation of all-sky weak lensing surveys.
  The disagreement is particularly pronounced in the squeezed limit.
\end{abstract}
\begin{keywords}: Cosmology-- Weak lensing -- Methods: analytical, statistical, numerical
\end{keywords}
%
\section{Introduction}
\label{sec:intro}
%

Thanks to recently completed Cosmic Microwave Background (CMB) experiments, 
such as the \textit{Planck} space mission\footnote{{\href{http://http://sci.esa.int/planck/}{\tt Planck}}}\citep{Planck,Planck18}, we now 
have a standard model of cosmology. 
There are however many outstanding questions that remain unanswered including, but not limited to, the nature of dark matter and dark energy;
or possible modifications of General Relativity (GR) on cosmological scales \citep{MG1,MG2,Planck2}. In addition the sum of the neutrino masses \citep{nu} remains
unknown. Not to mention aspects of the generation of the primordial fluctuations.
Fortunately it is expected that the weak lensing surveys in operation, including
Dark Energy Survey\footnote{\href{https://www.darkenergysurvey.org/}{\tt https://www.darkenergysurvey.org/}}(DES)\citep{DES}, Dark Energy Spectroscopic Instruments (DESI)\footnote{\href{http://desi.lbl.gov}{\tt http://desi.lbl.gov}},
Subaru Hyper-Suprime Cam
\footnote{\href{https://www.naoj.org/Observing/Instruments/HSC/index.html}
  {\tt https://www.naoj.org/Observing/Instruments/HSC/index.html}}, 
KiDS\citep{KIDS} and  near-future
CFHTLS\footnote{\href{http://www.cfht.hawai.edu/Sciences/CFHLS/}{\tt http://www.cfht.hawai.edu/Sciences/CFHLS}} as well as
Stage-IV large scale structure (LSS)
surveys such as \textit{Euclid}\footnote{\href{http://sci.esa.int/euclid/}{\tt http://sci.esa.int/euclid/}}\citep{Euclid},
the Large Synoptic Survey Telescope\footnote{\href{http://www.lsst.org/llst home.shtml}{\tt {http://www.lsst.org/llst home.shtml}}}\citep{LSST_Tyson} and
the Wide Field InfraRed Survey Telescope\citep{WFIRST}  will provide some clues or even answers to many of the questions that cosmology is facing.

Weak lensing (the minute distortions in the images of the distant galaxies by the intervening large-scale structure)
allows to extract information about the clustering of the intervening  mass distribution in the Universe \citep{review}.
The weak lensing surveys are complementary to spectroscopic galaxy redshift surveys such as
BOSS\footnote{\href{http://www.sdss3.org/surveys/boss.php}{\tt http://www.sdss3.org/surveys/boss.php}}\citep{SDSSIII}
or WiggleZ\footnote{\href{http://wigglez.swin.edu.au/}{\tt http://wigglez.swin.edu.au/}}\citep{WiggleZ} as they
provide an unbiased picture of the underlying dark matter distribution, whereas the galaxies and other biased 
tracers require a difficult modelling of the light to mass relationship \citep{bias_review} whose uncertainty is hard to assess precisely.

One challenge for weak lensing is that much of the information lies at small scales where clustering is non-linear
and non-Gaussian \citep{BernardeauReview}, and  theoretical predictions are challenging.
Another challenge is that the statistical estimates of cosmological parameters based on power spectrum 
analysis are typically degenerate,
e.g., $\sigma_8$ (the amplitude of mass density fluctuation) versus $\Omega_M$ (matter density parameter);
to overcome these degeneracies, external data sets e.g., CMB, and the addition of 
tomographic or 3D \citep{3D} information is often used. In order to address both of these challenges an alternative procedure
is to use higher-order statistics that probe the (quasi) non-linear regime
\citep{higher1, MunshiBarber292, higher2,higher3,MunshiBarber1,MunshiBarber3, Byun17, Song15}.

Previously it has been noted that even in the absence of any primordial non-Gaussianity \citep{Inflation},
gravitational clustering induces mode couplings that results in a secondary non-Gaussianity which
is more pronounced on smaller scales. Thus a
considerable amount of effort has been invested in understanding the gravity induced secondary non-Gaussianity
from weak lensing surveys. These statistics include the lower order cumulants \citep{MunshiJain1} and their correlators \citep{MunshiBias};
the multispectra including the skew-spectrum \citep{AlanBi} and kurtosis spectra \citep{AlanTri} 
as well as the entire PDF \citep{MunshiJain2} and
the statistics of hot and cold spots. Future surveys such as
\textit{Euclid} will be particularly interesting in this regard.
With its large fraction of sky-coverage it will be able to detect the gravity induced non-Gaussianity with a very
high signal-to-noise (S/N). It is also worth mentioning here that, in addition to breaking the degeneracy
between cosmological parameters, higher-order statistics are also
important for a better understanding of the covariance of lower-order estimators \citep{MunshiBarber3}.
The construction of morphological estimators such as the Minkowski Functionals are discussed in \citep{MF}.

The large fraction of sky covered with \emph{Euclid} will enable us to move from a real-sky correlation
function analysis to a more convenient spectral analysis in the harmonic domain. The methods developed
for analysing the CMB data will be helpful in this regard.
Several such estimators were developed and were used to analyse the
CMB maps generated by the \emph{Planck} satellite \citep{KSW, FLS, AlanBi, Bucher} and produced
the currently best available constraints on non-Gaussianity in inflationary models as well as
resulted in the detection of secondary non-Gaussianity \citep{Planck}.

In a recent paper a new set of estimators that are particularly sensitive to the
squeezed state of the bispectra also known as the {\em Integrated Bispectra} (IB) were proposed
\citep{IB}. These estimators
are interesting because of their simplicity, and ease of implementation.
Such estimators have also been used in 3D for quantifying galaxy clustering \citep{ChiangIB}
as well as in 1D to probe Lyman-$\alpha$ absorption features \citep{Lyman_bispec,chiang2}. 
Here, we go beyond the squeezed configuration and extend the results to
other shapes that include equilateral, folded and isosceles configurations.
We use the techniques developed in \citep{Namikawa18} for CMB lensing. Using numerical
ray-tracing simulations we check the validity of various approximations
in the construction of such estimators.

This paper is organized as follows. In \textsection{\ref{sec:theory}} we briefly summarize the theoretical results (for flat-sky as well as in all-sky limit).
The simulations are presented in \textsection{\ref{sec:simu}}.
The results are presented in \textsection\ref{sec:results}.
Finally, the discussions are presented in \textsection\ref{sec:disc}.

\section{Weak Lensing Convergence, Bispectrum and Estimators}
\label{sec:theory}
%
One of the first estimators developed for the study of non-Gaussianity is also know as the
KSW estimator \citep{KSW} which is a one-point estimator and is based on separable template fitting.
It is computationally fast but being an one-point estimator it can not differentiate independent contributions
to non-Gaussianity.

The skew-spectrum estimator developed in \citep{AlanBi} provides a power spectrum related
to the bispectrum. It is an extension of the one-point KSW estimator and
retains more information on the bispectrum. Being an $\ell$-dependent function
it can separate different scale-dependent contributions to the bispectrum and thus can be
valuable in separating systematics from true signal. An implementation
of skew-spectrum in the context of weak-lensing analysis will be presented
elsewhere (Munshi et al. 2019; in preparation).

The modal decomposition \citep{FLS} is a generalisation of KSW estimator and it relies on
the expansion of the target or theoretical bispectrum (template) as well as the bispectrum of
the map in a basis function. A general basis function can be tackled in this method
and separability is not a requirement.

\subsection{Angular Bispectrum}

\begin{figure}
  \centering
  \includegraphics[width=8.75cm]{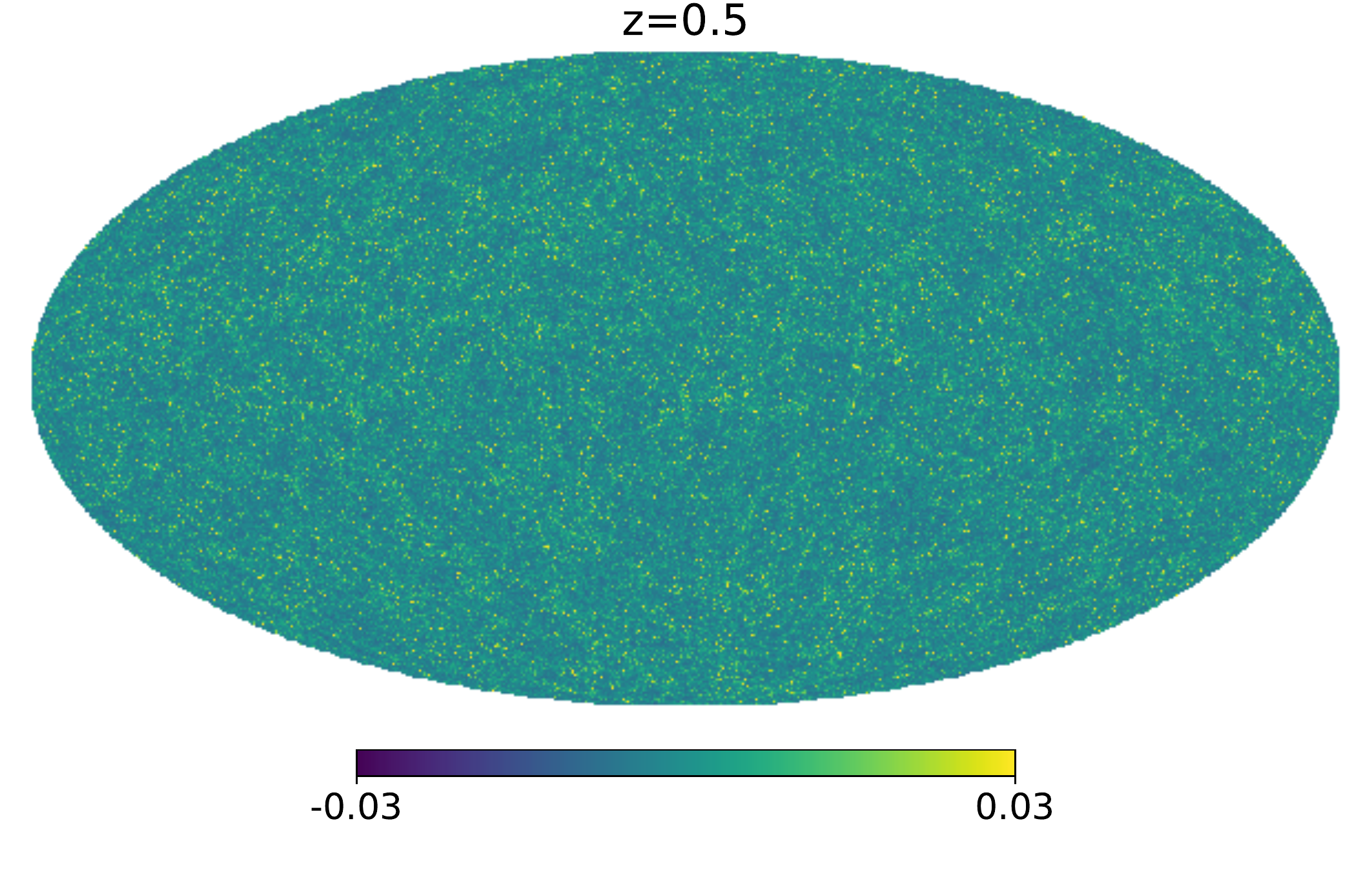}
   \includegraphics[width=8.75cm]{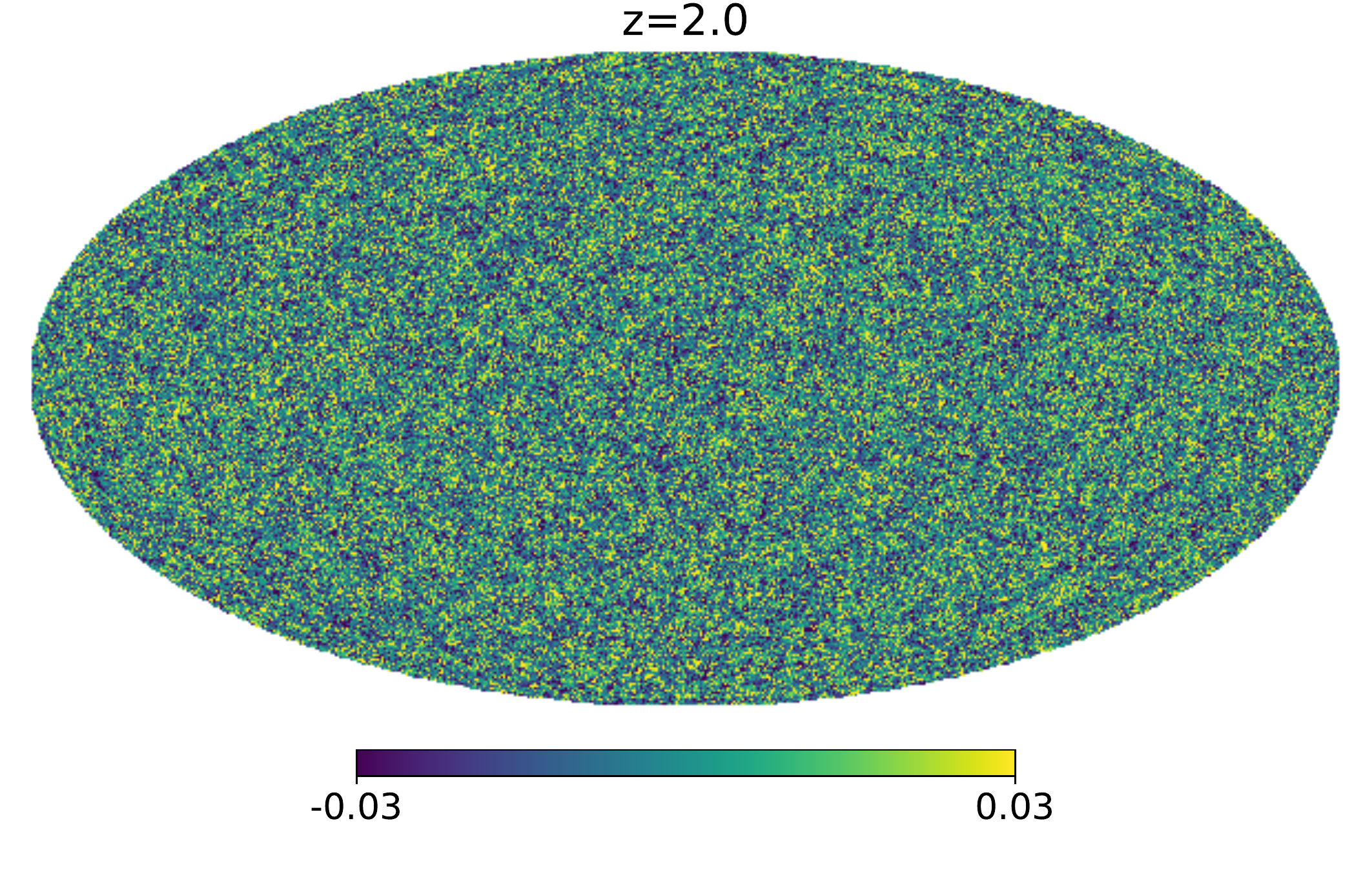}
  \caption{Examples of simulated $\kappa$ maps used in our study. The left panel corresponds to $z_s = 0.5$ while the right panel corresponds to
    $z_s = 2.0$. The maps were generated at a resolution of $N_{\rm side} = 4096$. See \textsection\ref{sec:simu} for more detail
    discussion about construction of maps used in our study.}
  \label{fig:eu}
\end{figure}
%
%
In this Section we provide a brief review of the existing literature and collate the available results
that will be later used in this paper. The weak lensing convergence \citep{review}, 
$\kappa^{\alpha}$ and its  
harmonic coefficients, $\kappa^{\alpha}_{\ell m}$, are related through the following
forward and reverse
harmonic transformations and are expressed in terms of the spherical harmonics of degree $\ell$ and order $m$: $Y_{\ell m}(\oh_p)$
defined over a discretized sphere as:
\ben
\kappa^{\alpha}_{\ell m} = \sum_{p} \kappa^{\alpha}(\oh_p) Y^{\star}_{\ell m}(\oh_p)\;\triangle \oh_p\;\; ; \quad
\kappa^{\alpha}(\oh_p) = \sum_{\ell m} \kappa^{\alpha}_{\ell m} Y_{\ell m}(\oh_p).
\label{eq:harmonics}
\een
The Greek index $\alpha$ labels tomographic bins. Here the  unit vector $\oh_p$ describes the angular position of the $p$-th pixel,
${}^*$ represents complex conjugation and $\triangle \oh_p$ denotes the pixel area (assumed to be the same for all pixels).
The three-point correlation function of the spherical harmonic coefficients
also known as the {\em angular bispectrum} is defined through the following expression;
see e.g. \cite{Inflation} for a review
\bes
\ben
&& \langle \kappa^{\alpha}_{\ell_1 m_1}\kappa^{\beta}_{\ell_2 m_2}\kappa^{\gamma}_{\ell_3 m_3}\rangle = {B}^{\alpha\beta\gamma}_{\ell_1\ell_2\ell_3}
 \left ( \begin{array}{ c c c }
     \ell_1 & \ell_2 & \ell_3 \\
     m_1 & m_2 & m_3
 \end{array} \right); 
\label{eq:bispectrum1}\\
&&    {B}^{\alpha\beta\gamma}_{\ell_1\ell_2\ell_3}
    = \sum_{m_1m_2m_3}
    \left ( \begin{array}{ c c c }
     \ell_1 & \ell_2 & \ell_3 \\
     m_1 & m_2 & m_3
    \end{array} \right)
    \langle \kappa^{\alpha}_{\ell_1 m_1}\kappa^{\beta}_{\ell_2 m_2}\kappa^{\gamma}_{\ell_3 m_3}\rangle.
    \label{eq:bispectrum}
\een
\ees
The angular brackets represent ensemble averaging. This form preserves the the rotational invariance of the three-point correlation
function in the harmonic domain.
 The quantity in parentheses is the Wigner 3j-symbol, which is nonzero only for triplets ($\ell_1, \ell_2, \ell_3$)
 satisfying the triangularity condition $|\ell_1-\ell_2| \le \ell_3 \le \ell_1+\ell_2$ and the condition
 that the sum $\ell_1 + \ell_2 + \ell_3$ is even, ensuring the parity invariance of the bispectrum (we will ignore any parity violating physics).
 Parity invariance imposes a selection rule that ensures underlying invariance of the field under
 {\em spatial inversion}. The analysis here can be extended to directly compute the bispectrum
 from shear $\gamma$ maps that would include both ``electric'' $\rm E$ and ``magnetic'' $\rm B$ modes. The parity violating contributions
 to bispectrum can be obtained by constructing a bispectrum that includes both $\rm E$ and $\rm B$ modes. This
 can be used to detect any possible parity violating physics as well as other systematic effects.
 
 The reduced bispectrum $b_{\ell_1\ell_2\ell_3}$ is typically defined through the following relation by introducing a geometric factor
 $h_{\ell_1\ell_2\ell_3}$ that only depends on $\ell$ and not on the convergence map $\kappa$:
\ben
b^{\alpha\beta\gamma}_{\ell_1\ell_2\ell_3} = h^{-1}_{\ell_1\ell_2\ell_3} B^{\alpha\beta\gamma}_{\ell_1\ell_2\ell_3}; \quad\quad
h^2_{l_1l_2l_3} = {\Sigma_{\ell_1}\Sigma_{\ell_2}\Sigma_{\ell_3}\over 4\pi}
\left ( \begin{array}{ c c c }
     \ell_1 & \ell_2 & \ell_3 \\
     0 & 0 & 0
\end{array} \right)^2 \, ;  \quad\quad \Sigma_{\ell_i} = 2\ell_i+1.
\label{eq:def_h}
\een
The reduced bispectrum $b_{\ell_1\ell_2\ell_3}$
was introduced by \cite{KS01}, see \cite{BCZ}
for an elaborate discussion, 
it is useful in establishing the all-sky and flat-sky correspondence. The factor $h_{\ell_1\ell_2\ell_3}$
represents the geometrical factors that depend on the harmonics $\ell$.
We will write Eq.(\ref{eq:bispectrum}) using the formal expression that will be used below to define the estimator:
\ben
{B}^{\alpha\beta\gamma}_{\ell_1\ell_2\ell_3} =
\left\la \int d\oh \, {\cal L}^{\alpha}_{\ell_1}(\oh) {\cal L}^{\beta}_{\ell_2}(\oh) {\cal L}^{\gamma}_{\ell_3}(\oh) \right\ra;
\quad\quad  {\cal L}^{\alpha}_{\ell_i}(\oh) = \sum^{\ell_i}_{m_i=-\ell_i} \kappa^{\alpha}_{\ell_i m_i} Y_{\ell_i m_i}(\oh).
\label{eq:construct_maps}
\een
The fields ${\cal L}^{\alpha}_{\ell_i}$ are constructed using maximally filtered convergence or $\kappa$
fields by keeping only one harmonic mode.
A binned version which we will implement is defined by constructing the ${\cal L}^{\alpha}_{b_i}$ which uses the coarse-grained
harmonics defined in a bin. Binning reduces the scatter in the estimator; however, unbinned
estimates can be used as well for parameter estimations. The binning will mix various shapes that we will discuss later. 
\ben
{B}^{\alpha\beta\gamma}_{b_1 b_2 b_3} =
\left\la \int d{\oh}\, {\cal L}^{\alpha}_{b_1}(\oh)\, {\cal L}^{\beta}_{b_2}(\oh) {\cal L}^{\gamma}_{b_3}(\oh) \right\ra \, ; \quad
         {\cal L}^{\alpha}_{{b}_i}(\oh) = \sum_{\ell \in b_i}\sum^{\ell}_{m =-\ell} \kappa^{\alpha}_{\ell m} Y_{\ell m}(\oh).
 \label{eq:binned_estimator}
 \een
 The angular brackets represent ensemble averaging.
 In the following, we will discuss the coarse-grained estimator for the reduced bispectrum.
 Notice that in addition to
 cross-correlating different tomographic bins, the estimator can also be used to cross-correlate external fields
 including e.g., the $y$-parameter maps from thermal Sunyaev-Zel'dovich (tSZ) effect.
 A simple expression for the covariance can be constructed in the noise dominated regime:
 \bes
 \ben
&& \langle\hat B^{\alpha\beta\gamma}_{\ell_1\ell_2\ell_3}\hat B^{\mu\nu\eta}_{\ell_4\ell_5\ell_6}\rangle
 =  h^2_{\ell_1\ell_2\ell_3} {\cal C}^{\alpha\mu}_{\ell_1}{\cal C}^{\beta\nu}_{\ell_2}{\cal C}^{\gamma\eta}_{\ell_3}
 \delta_{\ell_1\ell_4}\delta_{\ell_2\ell_5}\delta_{\ell_3\ell_6} + {\rm cyc.perm.}; \\
&&\sigma^2(B^{\alpha\beta\gamma}_{\ell_1\ell_2\ell_3}) = \sum_{\ell_1\in\Delta_1}\sum_{\ell_2\in\Delta_2}\sum_{\ell_3\in\Delta_3}
h^2_{\ell_1\ell_2\ell_3}  {\cal C}^{\alpha\beta}_{\ell_1}{\cal C}^{\beta\gamma}_{\ell_2}{\cal C}^{\gamma\alpha}_{\ell_3}  + {\rm cyc.perm.}
\een
\ees
The expressions ``cyc. perm.'' above represent a cyclic permutation of the superscripts (greek-indices)
which represents individual tomographic bins. The angular cross-spectra of two different bins
are represented by ${\cal C}^{\alpha\beta}_\ell=\la \kappa^{\alpha}_{\ell m}\kappa^{\beta *}_{\ell m}\ra$.
In case of a partial-sky coverage originating from an observational mask the linear term needs to be subtracted
to make the estimator unbiased \citep{AlanBi}:
\ben
B^{\alpha\beta\gamma\; \rm (lin)}_{\ell_1\ell_2\ell_3} =
\sum_p^{N_{\rm pix}} [{\cal L}^{\alpha}_{\ell_1}(\oh_p)\langle {\cal L}^{\beta}_{\ell_2}(\oh_p) {\cal L}^{\gamma}_{\ell_3}(\oh_p) \rangle
+  {\cal L}^{\beta}_{\ell_2}(\oh_p)\langle {\cal L}^{\alpha}_{\ell_1}(\oh_p) {\cal L}^{\gamma}_{\ell_3}(\oh_p) \rangle
+  {\cal L}^{\gamma}_{\ell_3}(\oh_p)\langle {\cal L}^{\alpha}_{\ell_1}(\oh_p) {\cal L}^\beta_{\ell_2}(\oh_p) \rangle].
\label{eq:lin}
\een
Here $N_{\rm pix}$ is the number of pixel and the discrete summation (which replaces integrals)
represents a pixelization scheme.
The ensemble averaging is computed by using many Gaussian realisations with the same power spectrum ${\cal C}_\ell$
and the resulting unbiased estimator $\hat B$ can now be constructed from the pseudo estimator $\tilde B$:
\bes
\ben
{\widehat B}^{\alpha\beta\gamma}_{\ell_1\ell_2\ell_3} = {\widetilde B}^{\alpha\beta\gamma}_{\ell_1\ell_2\ell_3} - B^{\alpha\beta\gamma \rm (lin)}_{\ell_1\ell_2\ell_3}.
\een
\ees
The pseudo estimator $\tilde B$ is constructed by restricting the integration in Eq.(\ref{eq:def_h}) and Eq.(\ref{eq:construct_maps})
to the observed region of the sky.
Here, ${\widehat B}^{\alpha\beta\gamma}_{\ell_1\ell_2\ell_3}$ represents the all-sky estimator which is constructed from a partial sky measurement
represented by ${\widetilde B}^{\alpha\beta\gamma}_{\ell_1\ell_2\ell_3}$ and corrections due to partial sky coverage is
encoded in the term  $B^{\alpha\beta\gamma \rm (lin)}_{\ell_1\ell_2\ell_3}$ defined in Eq.(\ref{eq:lin}). 
%
\subsection{Estimator}
\label{sec:bin}
%
Using the definitions in the previous section we outline the estimator here that will be used for analysing the simulations.
The reduced bispectrum $b_{\ell_1\ell_2\ell_3}$ can be estimated by the following expression (see \citep{Namikawa18} for a detailed derivation).
\ben
b_{\ell_1\ell_2\ell_3} =
{1 \over h^2_{\ell_1\ell_2\ell_3}} \sum_{p}^{N_{\rm pix}} {\cal L}^{\alpha}_1(\oh_p){\cal L}^{\beta}_2(\oh_p){\cal L}^{\gamma}_3(\oh_p) \approx
\sqrt{4\pi}{\sum^{N_{\rm pix}}_p {\cal L}^{\alpha}_{\ell_1}(\oh_p){\cal L}^{\beta}_{\ell_2}(\oh_p){\cal L}^{\gamma}_{\ell_3}(\oh_p) \over \sum^{N_{\rm pix}}_p h_{\ell_1}(\oh_p)h_{\ell_2}(\oh_p)h_{\ell_3}(\oh_p) }
\label{eq:estimator}
\een
The first equality uses the expression of $b^{\alpha\beta\gamma}_{\ell_1\ell_2\ell_3}$ in terms of $B^{\alpha\beta\gamma}_{\ell_1\ell_2\ell_3}$  as given in Eq.(\ref{eq:def_h})
and uses the definition of ${\cal L}^{\alpha}_{\ell_i}$ in Eq.(\ref{eq:construct_maps}) to express $B^{\alpha\beta\gamma}_{\ell_1\ell_2\ell_3}$.
The individual $\cal{L}$ maps are
constructed using modes estimated from maps as described in Eq.(\ref{eq:binned_estimator}).
The construction of the denominator can be done either using the
standard textbook \citep{Edmonds68} expression or using similar technique used for the numerator. 
We start by noticing the fact that $h^2_{\ell_1\ell_2\ell_3}$ defined in Eq.(\ref{eq:def_h}) can be expressed as a Gaunt Integral as follows:
\bes
\ben
 h^2_{\ell_1\ell_2\ell_3}  =  \sqrt{(2\ell_1+1)(2\ell_2+1)(2\ell_3+1)\over 4\pi} \int Y_{\ell_1 0}(\oh)Y_{\ell_2 0}(\oh)Y_{\ell_3 0}(\oh)d\oh.
\een
 Next, we define the functions $h_i(\oh_p)$ as follows and use a discrete sum to replace the integral: 
\ben
h^2_{\ell_1\ell_2\ell_3}   ={1\over\sqrt{4\pi}} \sum^{N_{\rm pix}}_p{h_1(\oh_p)}{h_2(\oh_p)}{h_3(\oh_p)}; \quad h_{\ell_i}(\oh_p)=
\sum_{\ell} \sqrt{(2\ell+1)}  Y_{\ell_i 0}(\oh_p).
\een
\ees
This factor does not depend on the source-redshift or convergence power spectrum and has a purely geometric origin.
To reduce the computational cost as well as the scatter we provide the bin estimates instead of the mode-by-mode
result. Notice that binning mixes different types of bispectrum as all possible modes in a given bin
contribute. However, this effect can be modelled theoretically before comparing with data. Reducing the
size of the bins can effectively control such contamination.
Previous study by \cite{Namikawa18} has found the effect of binning
to be significant for all shapes but comparatively less significant for the equilateral case.
The binning can be introduced through the following expression:
\ben
b^{\alpha\beta\gamma}_{b_1b_2b_3} =
{\sum_{\ell_i\in b_i} h^2_{\ell_1\ell_2\ell_3} b^{\alpha\beta\gamma}_{\ell_1\ell_2\ell_3} \over
  \sum_{\ell_i\in b_i} h^2_{\ell_1\ell_2\ell_3}}.
\label{eq:binned}
\een
As we mentioned before, the above estimator can also be generalised to analyse directly the shear maps by decomposing them
into {\em electric} E and {\em magnetic} B modes. The linear terms
appearing due to partial sky coverage will be introduced later in this section.

Typically two different methods are pursued in the detection of non-Gaussianity. The first method
uses a template for the detection relying on a matched-filter technique. In this case, the spherical harmonics coefficients are convolved with
weights to make them {\em near optimal}. Such methods work in the weakly non-Gaussian
case. In case of lensing, non-Gaussianity is more dominant and optimality is less of an issue as signal-to-noise is rather high.
We choose a non-parametric estimator (also called a binned estimator) which is sub-optimal but that does not rely on a theoretical model. 
\begin{figure}
  \centering
  \includegraphics[width=6cm]{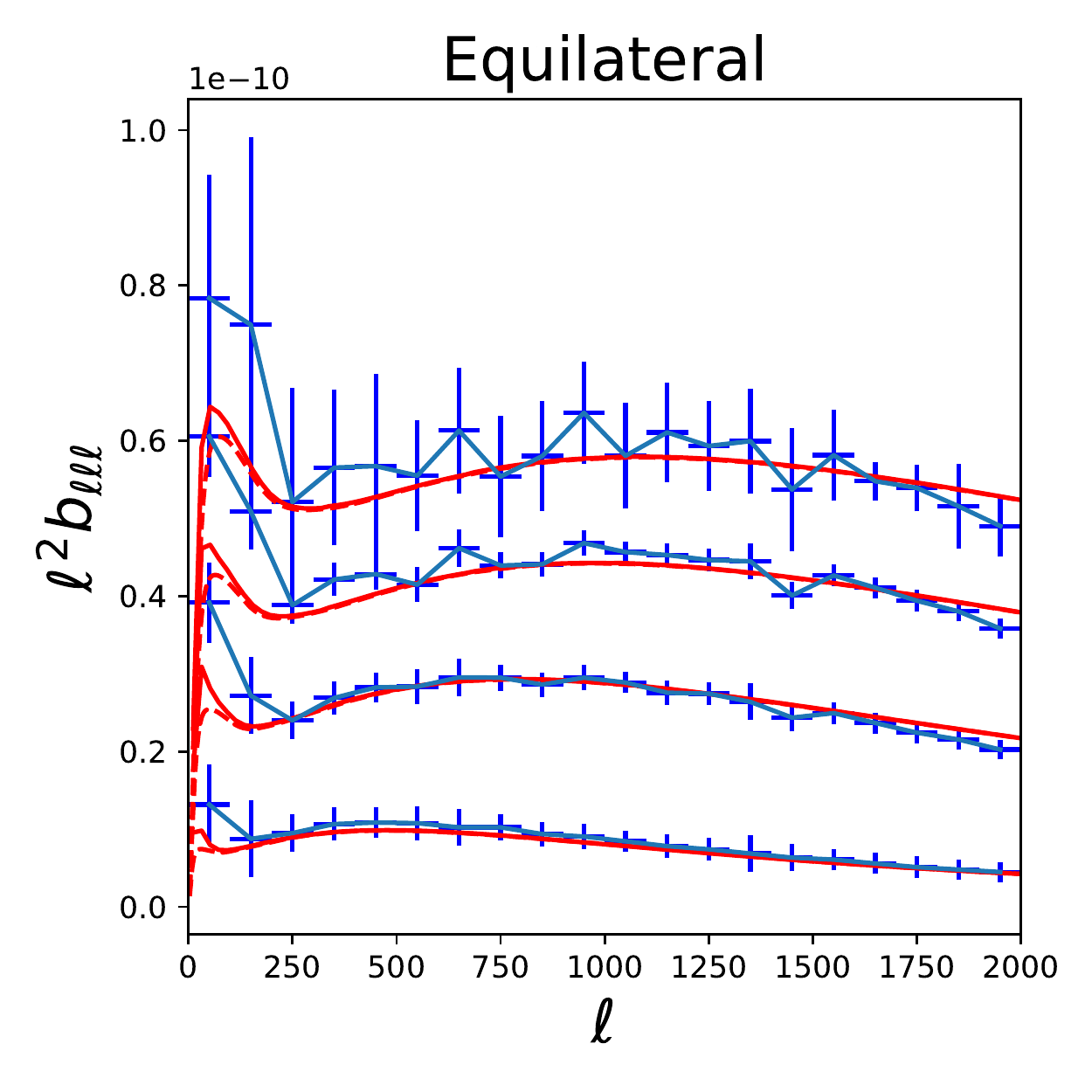}
  \includegraphics[width=6cm]{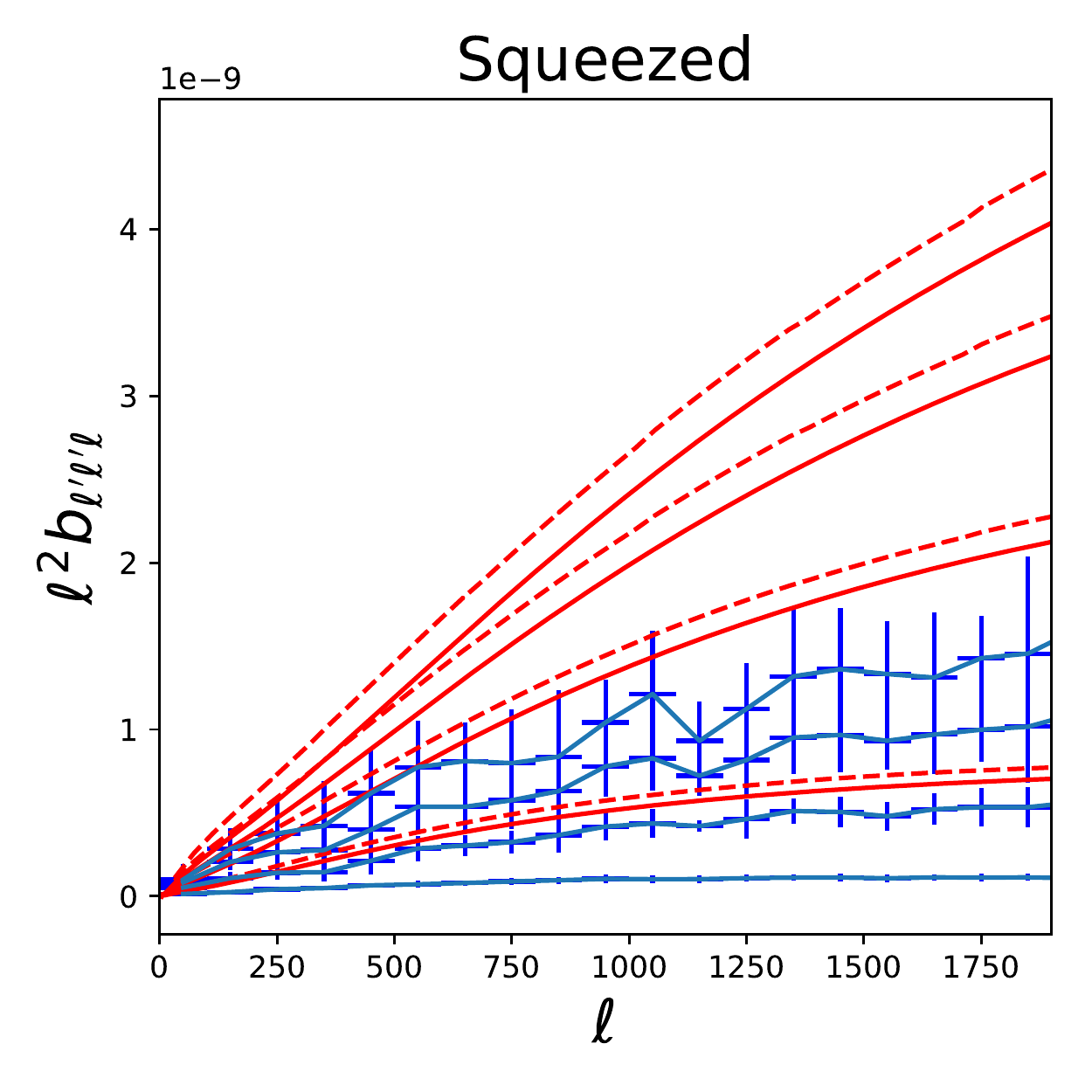}\\
  \includegraphics[width=6cm]{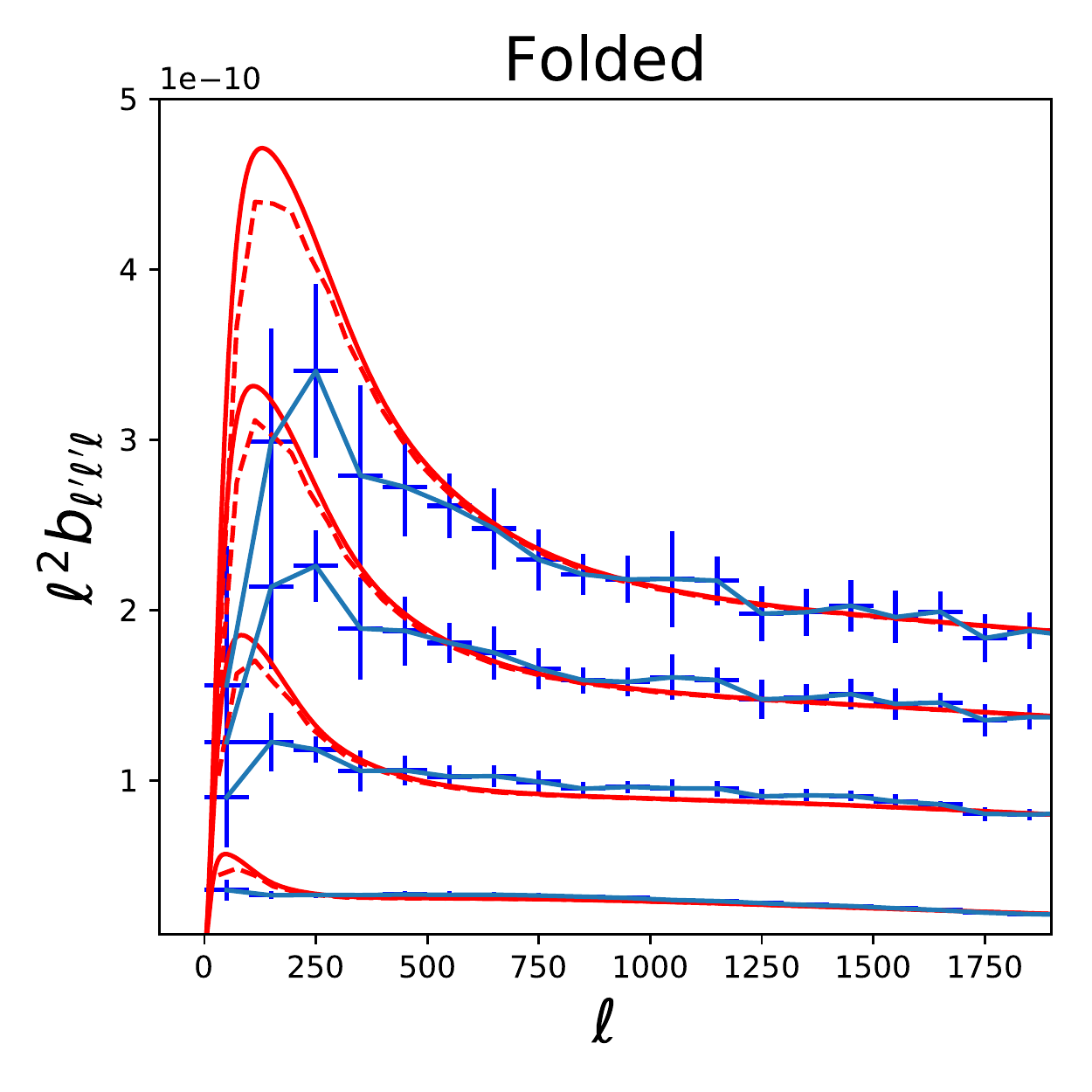}
  \includegraphics[width=6cm]{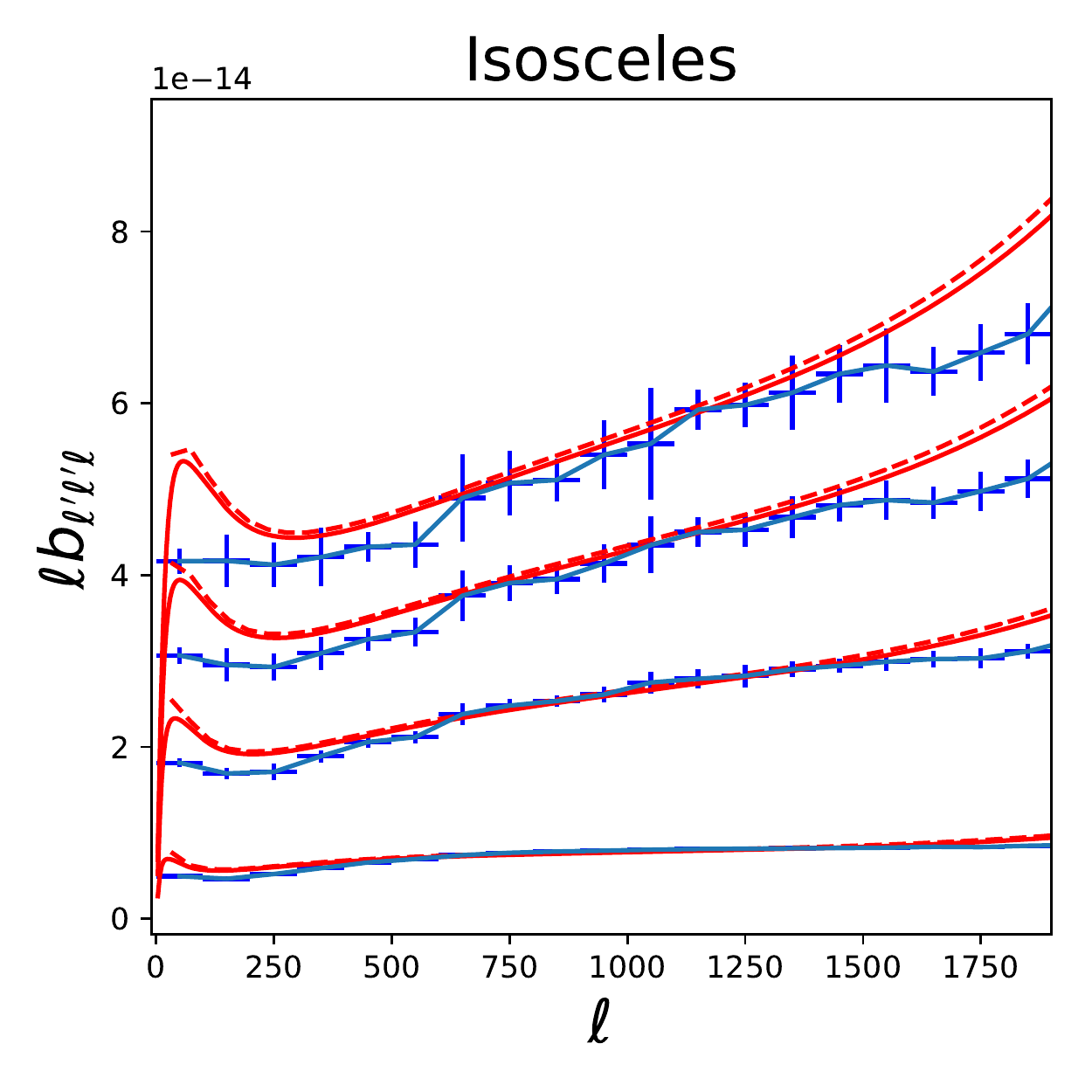}
  \caption{The reduced bispectrum $b_{\ell_1\ell_2\ell_3}$ measured in simulations for four different shapes are compared against theoretical predictions.
    The upper left-panel
     corresponds to the {\it equilateral} configuration $\ell_1=\ell_2=\ell_3=\ell$ and depicts
     $\ell^2 b_{\ell\ell\ell}$ as a function of $\ell$. The sources are placed
     at different redshifts; from bottom to top they correspond respectively to $z_s=0.5,1.0,1.5$ and $2.0$.
     The theoretical predictions are computed using the fitting function given in \citep{GilMarin}.
    The maps used for this analysis were constructed at $N_{\rm side}=4096$. No noise
    or mask were included.  For the {\it squeezed case} (upper right-panel) the bispectrum  $b_{\ell'\ell'\ell}$ is shown for $\ell_1=\ell'=50$
    and $\ell_2=\ell_3=\ell$.
    The error-bars were computed using ten realisations for each redshift. We use a total of 20 bins to
    cover the range of multipoles. The solid lines in each panel correspond to theoretical predictions computed using the
    fit in Eq.(\ref{eq:bi1}-\ref{eq:bi3b}). The dashed lines include the effect of binning. The expressions for the
    binned bispectrum are presented in Eq.(\ref{eq:binned}).  The lower left-panel shows the {\it folded} configuration while the lower right-panel
    depicts the {\it isosceles} configuration.
    For the isosceles case we take $\ell_1=\ell$ and $\ell_2=\ell_3=1000$.
    For the folded case we have $\ell_1=2\ell_2=2\ell_3=\ell$ and we plot $\ell b_{\ell'\ell'\ell}$ as a function of $\ell$.
   Our results are derived using flat-sky approximations.}
  \label{fig:all}
\end{figure}
%
\subsection{Convergence Bispectrum and Matter Bispectrum}
%
In this section we express the weak lensing bispectrum ${\cal B}$ as a line of sight integration
of bispectrum $B_{\delta}$ associated with the density contrast $\delta$.
The gravitational deflection potential $\phi$, which is related to the convergence $\kappa$,
can be expressed as a line of sight integration of Weyl potential $\Phi$ as follows:
\bes
\ben
&& \phi(\oh) = -2 \int_0^{\chi_s}  d\chi W(\chi,\chi_s)\Phi(\chi,\oh); \quad\quad
W(\chi,\chi_s) = {\chi_s -\chi \over \chi\chi_s} \Theta(\chi_s -\chi); \quad\quad \kappa_{\ell m} = {1\over 2}\ell(\ell+1)\phi_{\ell m}.
\een
\ees
Here, $W(\chi,\chi_{\star})$ represents the lensing kernel and $\chi$ is the radial comoving distance and
$\Theta$ is the unit step function.
The comoving radial distance to the source plane is denoted as $\chi_s$. Using the {\em flat-sky} approximation
the convergence bispectrum ${\cal B}_{\ell_1\ell_2\ell_3}$ can be related to the underlying
matter bispectrum $B_{\delta}$ through the following line-of-sight integration:
\ben
    {\cal B}_{\ell_1\ell_2\ell_3} = \int_0^{\chi_s}
    d\chi \left [ {3\Omega_{\rm M}} H_0^2 \over 2a(\chi) \right ]^3
    \chi^2 W^3(\chi,\chi_s) B_{\delta}\left ({\ell_1\over \chi},{\ell_2\over \chi},{\ell_3\over \chi},\chi \right ); \quad k_i ={\ell_i \over \chi}.
    \label{eq:define_kappa_bispec}
\een
The flat-sky bispectrum is defined in terms of the 2D Fourier coefficients $\kappa(\bl)$:
\ben
\langle \kappa(\bl_1)\kappa(\bl_2)\kappa(\bl_3) \rangle = (2\pi)^2 \delta_{\rm 2D}({\bl}_1+{\bl}_2+{\bl}_3)b(\bl_1,\bl_2,\bl_3).
\een
The reduced bispectrum satisfies the relationship $b_{\ell_1\ell_2\ell_3} \approx b(\bl_1,\bl_2,\bl_3)$ \citep{Inflation}.
Our analytical predictions are derived using the flat-sky approximation, though, the numerical results are derived
using all-sky simulation we expect that for $\ell > 100$ these two results should match; see e.g. \cite{small_angle} for related discussions
in the context of power spectrum. The gravity induced matter bispectrum $B_{\delta}(\bk_1,\bk_2,\bk_3,\chi)$ is often modelled 
using standard tree-level perturbation theory in the quasilinear regime. 
We use a fitting function for $B_{\delta}$ in our computation of ${\cal B}$ using Eq.(\ref{eq:define_kappa_bispec})
which is described in Appendix-\textsection\ref{sec:append1}. A generalization to 3D bispectrum
is presented in in Appendix-\textsection\ref{sec:append2}.
%
\subsection{Shapes of Weak Lensing Convergence Bispectrum}
The bispectrum represents three-way interaction of three different wave vectors in 3D or harmonic multipoles in spherical
sky. To guarantee translational invariance in the case of 3D and rotational invariance on the sphere
the vectors should form a closed triangle. Various triangular configurations have been
studied in great detail in the context of inflationary models and they can be related to a specific
inflationary mechanism, e.g., the {\em squeezed model} is related to
the {\em local} form of non-Gaussianity \citep{SBa,SBb,Gangui}, and the {\em equilateral}
form of primordial non-Gaussianity appears in the inflationary models with
a non-canonical kinetic term \citep{Ali04, ChenEasther07,Cheung08,cremi06}.

Concerning the input of large scale structure formation scenarios, the various shapes can also give
clues on the gravitational dynamics at work to create the resulting
convergence maps.
For the equilateral case the length of three sides of the triangle are equal and
in real space  this bispectrum can represent possibilities as dictated by the signature
of the bispectrum. A positive amplitude for the reduced bispectrum will correspond to
compact collapsed over dense regions surrounded by underdense regions. Whereas the
negative amplitude corresponds to underdense regions surrounded by over-dense
objects. The positive bispectrum will correspond to positive skewness (representing long tail of
positive $\kappa$ values) and similarly a negative bispectrum will represent negative
skewness that correspond to long tail of negative $\kappa$ values in probability distribution
function of $\kappa$. The formation of large scale structure formation  is known to induce a positive skewness
and hence a positive amplitude for the reduced bispectrum.

Exact equilateral configurations are less likely just as numerical simulations
in real space do not often produce exact spherically collapsed over-dense regions.
The elliptical or elongated over-dense regions are represented by folded
configuration of bispectrum. In the limiting case these can take the shape
of filaments. 

The squeezed configuration corresponds to the situation when one wave vector
is much smaller compared to the other two. In the extreme squeezed limit when one of the
wave vectors approaches zero, the other two wave vectors are equal and opposite to
each other. The squeezed configuration thus can be understood
as a {\em response} to a long wavelength modulation of the local power spectrum.
This response function approach \citep{Response17} have recently been studied and investigated
in detail by many authors in a scenario also known as the separate universe
formalism \citep{Baldauf,LHT14}. This method was recently studied in the context of weak lensing
convergence maps using a different approach that we use here  by \citep{IB}.
For a Gaussian random field, the coupling of local fluctuations to
long wavelength modulation is zero. For a positive bispectrum,
long term fluctuations enhance the small scale
collapse. The squeezed limit has also been studied in diverse cosmological
context, e.g., in the case of CMB lensing, galaxy clustering as well as
Lyman-$\alpha$ absorption studies; see e.g. \citep{ChiangIB, chiang2, Lyman_bispec, IB19} and references therein.
The study of squeezed limits are important from another aspect. In recent
studies consistency relations were obtained in the squeezed limit
which can be used as a diagnostics to test any violation from
Gaussian initial conditions or modification of gravity. The test of such
consistency relations can be performed using weak lensing observations
that can give a rare glimpse of non-perturbative regime of gravitational clustering \citep{KehagiasRiotto}.

Clearly the one-to-one correspondence of 3D shapes and their 2D counterparts
is lost due to projection. 
The numerical results for these configurations will be presented in \textsection\ref{sec:disc}.
%

\section{Simulations}
\label{sec:simu}
%
%
We use the all-sky weak-lensing maps generated by \citep{Ryuchi}\footnote{http://cosmo.phys.hirosaki-u.ac.jp/takahasi/allsky\_raytracing/}
These maps were generated using ray-tracing through N-body simulations using
multiple lens planes and to generate convergence $\kappa$ as well as shear $\gamma$ maps.
The source redshifts used were in the range $z_s= 0.05-5.30$ at an interval $\Delta \chi=150 h^{-1}$ Mpc
(corresponding to $\Delta z_s=0.05$ at the nearby universe).
The simulations also include maps generated for sources placed at last scattering surface
for CMB lensing studies. The simulations do not assume the Born approximations
and do include post-Born corrections and its higher-order analogs to an arbitrary order.
However, such corrections are expected to be negligible for the low source redshift 
that we are interested here \citep{Namikawa18}. Many checks were conducted using these maps
and were found to be in good agreement with theoretical predictions.
For a given redshift, sets of maps were generated using different resolution
in {\tt HEALPix}\footnote{https://healpix.jpl.nasa.gov/} format \citep{Gorski} which uses equal area pixelization scheme.
The number of pixels $N_{\rm pix} = 12 N^2_{\rm side}$ is defined in terms of the resolution parameter
$N_{\rm side}$. Detail test of electric/magnetic (E/B) decomposition of
shear maps and construction of $\kappa$ maps can be found in \citep{Ryuchi}.
For our purpose we use the $\kappa$ maps generated at $N_{\rm side}=4096$.
These maps have been cross checked using higher resolution maps constructed
at $N_{\rm side}=8192, 16384$ and were found to be in good agreement with theoretical predictions
up to the angular harmonics $\ell \le 3600$. The corresponding angular resolutions are $0.86'$, $0.43'$ and $0.21'$
for $N_{\rm side} =4096, 8192$ and $16384$ respectively. In our study we will be restricting
us to $\ell \le 2048$. Beyond this limit it is expected that baryonic feedback may play
increasingly dominant role \citep{Weiss19}. These maps were recently used to analyse the bispectrum
in a recent publication but in the context of CMB lensing \citep{Namikawa18}. In Figure-\ref{fig:eu} we show examples of
maps used in our study.

These simulations assume a 
$\Lambda$CDM background cosmology with the following set of cosmological parameters $\Omega_{\rm CDM} = 0.233$, $\Omega_b = 0.046$ and
$\Omega_{\rm M} = \Omega_{\rm CDM}+\Omega_b = 0.279$. The Hubble parameter was assumed to be $h=0.7$, the amplitude of
density flcutuation $\sigma_8=0.82$ and the spectral index $n_s=0.97$. 
%
\section{Results}
\label{sec:results}
%
Bispectrum and other higher-order multispectra will be detected in ongoing and future weak lensing surveys with
a high S/N ratio. In this paper we compare the theoretical modelling of bispectrum
with high resolution all-sky numerical simulations. In principle all possible configurations
of the bispectrum should be considered. This is computationally demanding so, we have
concentrated on specific configurations of the bispectrum.
In our numerical investigation we have considered the following representative shapes for the bispectrum:
(A) Squeezed: $\ell_1=50,\ell_2=\ell_3=\ell$ \quad (B) Equilateral: $\ell_1=\ell_2=\ell_3=\ell$ 
(C) Isoceles: $\ell_1=\ell$, $\ell_2=\ell_3=1000$ and \quad  (D) Folded: $\ell_1=2\ell_2=2\ell_3=\ell$.
 We have restricted our comparison to $\ell<2048$ modes as it is expected that the
  higher-order multipoles will be affected by baryonic feedback that are difficult to model accurately. For each model
  we have considered four tomogrphaic bins $z_s=0.5,1.0,1.5,2.0$. The source redshift $z_s$ increases from bottom to top
  in each panel for various curves shown in Figure-\ref{fig:all}.

The results for equilateral configurations are shown in Figure-\ref{fig:all} (upper left-panel).
The numerical results
  agree reasonably well with the theoretical predictions based on commonly used fitting functions for $\ell > 200$, 
  at least for $\ell < 2048$. This is consistent with the general trend of results reported in \citep{Namikawa18}.
  This deviation of numerical results at lower $\ell$s from theoretical predictions is visible in all of the models we have studied and is
  likely to be related to
  the use of flat-sky approximation in our theoretical derivations. Corrections,
  to flat-sky results can be included in a systematic manner \citep{small_angle}.
  The results of comparison for the squeezed model is presented in (upper right-panel).
  We see maximum departure from theoretical predictions in case of squeezed configuration
  where the theoretical predictions can be higher by a factor of two compared
  to the results obtained from simulations. 
  The Figure-\ref{fig:all} depicts the folded (lower left-panel) and the isosceles (lower right-panel) case.
  In case of the folded model we find significant departure from theoretical predictions at
  $\ell < 500$. The theoretical model can predict the results from simulations reasonably well.
  However, the small $\ell$ discrepancy is far more pronounced than other models.
  The isosceles case only matches with theoretical predictions for $\ell=1024$ where the
  configuration is actually close to the equilateral case.

  \cb{While corrections to the flat-sky approximation may explain suppression of the low $\ell$ results
      in case of equilateral
  model, the finite volume corrections can not be ignored for other cases. The bispectrum is a nonlinear
  statistic. The effect of finite volume is well studied in the real-space
  at the level of one point statistics, e.g., cumulants and their correlators. These effects are also
  well documented
  for estimation of power spectrum also known as the integral constraint - see e.g. \cite{BernardeauReview}.
  However, similar studies are lacking for the bispectrum which is proportional
  to the product of two power spectra probing different scales
  and hence are expected to be more affected by such an effect.
  This is also true in real space where e.g. skewness and other higher-order moments are more
  affected by the finite volume effect than the variance. The finite volume
  effect is more pronounced in the highly nonlinear regime. In equilateral case only one scale is
  involved making its interpretation comparatively simpler. For other three cases, different
  levels of nonlinearities are probed by the different wave numbers of the bispectrum.
  In any case, the fact that finite volume correction is a factor in our results, can easily be seen
  from Figure-\ref{fig:all}.
  The CMB lensing on the other hand,
  where the source is at a rather high redshift $z_s\approx 1100$, probes the quasi-linear
  regime thus reducing the effect of finite volume corrections. The corrections to the
  flat-sky expressions are also less prominent for a high source redshift. A broader binning can reduce
  the discrepancy between theory and simulations that we have noted but at the cost of degrading the
  cosmological information content of the spectrum.}.

  \cb{The accuracy of the fitting function we use may not be adequate to describe the range of redshift and
    harmonic modes we have studied. An improved fitting function for the bispectrum, that is currently
  underway will be presented elsewhere (Takahashi et al. 2019, in preparation).}
 
\section{Discussion}
\label{sec:disc}

The gravity induced bispectrum, arising from non-Gaussianity in weak lensing maps, can help to tighten
the constraints on cosmological parameters. We have tested available theoretical fitting functions
accuracy to reproduce numerical results. The lensing bispectrum as a possible probe for modified gravity theories as well as
neutrino masses will be a natural future step.

Impact of modified gravity theories on CMB lensing induced bispectrum was studied
in \citep{Namikawa18} to some extent.
Theoretical modelling of bispectrum in modified gravity scenarios is more challenging.
  Extensions of the perturbative approach was considered in
  \citep{Bose1,Bose2} by introducing more freedom to the kernels and validating them using
  numerical simulations. Such approaches and
  others including halo model predictions have already been proposed with
  varying degree of success for two main type of modified gravity theories,
  i.e., (A) models with Vainshtein-screening mechanism which includes the DGP model
  as well as the Hordenski \citep{Hordensky74} and beyond Hordenski theories
  \citep{BeyondHordensky1,BeyondHordensky2, BeyondHordensky3}
  and (B) the models with Chameleon-screening that includes the Hu-Sawicki $f({\rm R})$ model \citep{fofR}.
  In the DGP model \citep{DGP} the bispectrum from simulations can be reproduced using
  the GR expression by suitably modifying the power spectrum.
  The situation is somewhat more complicated for $f(R)$ theories.
  However, the fitting used for GR needs first to be standardised more accurately.
  Only then can the deviations from them can be studied meaningfully. The numerical modelling
  is more important at small scales where analytical results start to fail.
  In future it will also be interesting to consider the effect of neutrino mass
  on bispectrum when simulated all-sky lensing maps for such cosmologies will be available \citep{Liu17,Coulton18}.
  
  In addition to studying various shapes of weak lensing bispectrum
  we have extended the framework in several directions. We have computed the
  covariance of our estimator in the noise dominated regime in the weakly non-Gaussian limit.
  We have shown how the estimator can include tomographic bins.
  Using spherical Fourier-Bessel (sFB) decomposition,  we have extended the estimator to 3D.
  Inclusion of weights to make the estimator near-optimal is described in the limit of vanishing
  non-Gaussianity, i.e., in the noise dominated regime. We also provide a simplistic estimate of the covariance.
  The extension to next higher-order namely to the level of the reduced trispectrum is outlined in appendix-\ref{sec:append_tri}.
%

%
\section*{Acknowledgment}
DM is supported by a grant from the Leverhume Trust at MSSL.
TDK is supported by Royal Society University Research Fellowship.
RT is supported by MEXT/JSPS KAKENHI Grant No. JP15H05893 and 17H01131.
BB acknowledges support from the Swiss National Science Foundation (SNSF) Professorship grant No. 170547.
\appendix
\section{Fitting Functions to The Matter Bispectrum}
\label{sec:append1}
The fitting functions were provided by \citep{SC,GilMarin}. They have been tested in many different contexts.
In a recent paper \citep{Namikawa18} they were used to model CMB lensing. Following earlier studies by various authors \citep{TJ04,KS05,Coulton18},
we use these functions to model weak lensing at low source redshifts. This is important given
that the ongoing and future mission will be able to measure the lensing-induced bispectrum
with high level of significance. This requires not only a (near) optimal data analysis
pipeline but also an accurate theoretical modelling.

The perturbative matter bispectrum has the following form in terms of the perturbative kernels $F_2$ \citep{BernardeauReview}:
\ben
&& B_{\delta}(\bk_1,\bk_2,\bk_3; \chi)
= 2F_2(\bk_1,\bk_2, z) P_{\delta}(\bk_1,z)P_{\delta}(\bk_2,z)
+\rm{cyc.perm.}
\label{eq:bi1}
\een
where $P_{\delta}(\bk,z)$ is the matter power spectrum at a redshift $z$. The  second order kernel $F_2$ can be computed
using Eulerian perturbation theory which has the following form with $a=b=c=1$. However, comparison against numerical
simulation shows that phenomenological extension is possible by making the paramaters $a$, $b$ and $c$ scale and redshift
dependent \citep{SC, GilMarin}.
\ben
&& F_2(\bk_1,\bk_2,z) = {5\over 7} a(k_1,z)a(k_2,z) 
+ {1 \over 2}{k_1^2 + k_2^2 \over k_1k_2} b(k_1,z)b(k_2,z) \cos\theta
+ {2 \over 7} c(k_1,z)c(k_2,z).
\label{eq:bi2}
\een
We list the functions that appear in the fitting function for completeness:
\bes\ben
\label{eq:bi3a}
&& a(k,z) = {{1+ {\sigma_8(z)}}^{a_6}\sqrt{0.7 Q(n_{\rm eff})} (q a_1)^{n_{eff}+a_2}
\over 1+(qa_1)^{n_{eff}+a_2}} \\
&& b(k,z) = {1+ 0.2a_3(n_{\rm eff}+3)(q a_7)^{n_{\rm eff}+3+a_8}
\over 1+(qa_1)^{n_{\rm eff}+3.5+a_8}} \\
&& c(k,z) = {1+ [4.5a_4/(1.5+(n_{\rm eff}+3)^4)](qa_5)^{n_{\rm eff}+3 +a_9}
  \over 1+ (qa_5)^{n_{\rm eff}+3.5 + a_9}}.
\label{eq:bi3b}
\een\ees
The amplitude of the bispectrum is fixed by the amplitude $Q(x) = (4-2^x)/(1+2^{x+1})$ and $q = {k/k_{\rm NL}}$ with $k_{\rm NL}$ defined through the
following equation $4\pi k^3_{\rm NL} P^{\rm lin}(k_{\rm NL})=1$ and $n_{\rm eff}= d\ln P^{\rm lin}(k)/d\ln k$.
Two different fitting functions are available in the literature.
An initial fitting function was suggested by \cite{SC}.
An improved fitting function was given by the following parameters in \cite{GilMarin} which we have used:
\ben
a_1 =0.484, a_2=3.74, a_3=-0.849, a_4=0.392, a_5=1.01, a_6=-0.575, a_7=0.128, a_8=-0.722, a_9=-0.926.
\een
It is expected that some modified gravity theories would also leave a detectable signature
in convergence or $\kappa$ maps. However, the theoretical modelling requires an accurate
framework. At present there are no fitting functions for modified gravity theories.
Various  heuristics schemes from perturbative calculations to phenomenological
modification of Eq.(\ref{eq:bi1}) have been developed recently \citep{Bose1,Bose2}.
\section{Weak Lensing Bispectrum in 3D}
\label{sec:append2}
With the advent of accurate photometric redshift information a full 3D analysis beyond tomographic
binning has become possible. 
A 3D decomposition of the convergence maps can be accomplished using a spherical Fourier-Bessel transform following basis function.
The forward and reverse transforms have the following form \citep{3D,higher3}:
\ben
\kappa_{\ell m}(k) = \sqrt{2\over \pi}\, \int d^3{\bf r}\, \kappa({\bf r})\, j_{\ell}(kr)\, Y^*_{\ell m}(\oh); \quad\quad
\kappa(\br) = \sqrt{2\over \pi}\, \int\, dk\, \sum_{\ell}\sum_{m=-\ell}^{\ell}\, \kappa_{\ell m}({k})\,  j_{\ell}(kr)\, Y_{\ell m}(\oh).
\een
The 3D basis functions are constructed using the eigenfunctions of the Laplacian operator taking advantage of
the radial symmetry $j_{\ell}$ above corresponds to the spherical Bessel function and $k$ corresponds to the
radial wave number and $r = |{\bf r}|$ is the radial co-moving distance.
The corresponding 2D bispectrum defined in Eq.(\ref{eq:bispectrum1})-Eq.(\ref{eq:bispectrum}) now gets generalised to:
\bes
\ben
&& \langle \kappa_{\ell_1m_1}(k_1) \kappa_{\ell_2m_2}(k_2)\kappa_{\ell_3m_3}(k_3)\rangle = \left ( \begin{array}{ c c c }
     \ell_1 & \ell_2 & \ell_3 \\
     m_1 & m_2 & m_3
\end{array} \right){\cal B}_{\ell_1\ell_2\ell_3}(k_1,k_2,k_3)\\
\label{eq:3Dbi1}
&&        {\cal B}_{\ell_1\ell_2\ell_3}(k_1,k_2,k_3) =
       \sum_{m_1m_2m_3} \left ( \begin{array}{ c c c }
     \ell_1 & \ell_2 & \ell_3 \\
     m_1 & m_2 & m_3
       \end{array} \right)\langle \kappa_{\ell_1m_1}(k_1) \kappa_{\ell_2m_2}(k_2)\kappa_{\ell_3m_3}(k_3)\rangle.
       \label{eq:3Dbi2}
       \een
       \ees
       Following the same series of arguments the estimator presented in Eq.(\ref{eq:binned_estimator}) now gets generalised to:
\ben
{\cal B}^{}_{\ell_1\ell_2\ell_3}(k_1,k_2,k_3) =
\left\la \int d\Omega {\cal L}^{}_{1}(\oh) {\cal L}^{}_{2}(\oh) {\cal L}^{}_{3}(\oh) \right\ra;
\quad\quad  {\cal L}^{}_{i}(\oh) = \sum^{\ell_i}_{m_i=-\ell_i} \kappa^{}_{\ell_i m_i}(k_i) Y_{\ell_i m_i}(\oh).
\een
A binning in $k$ can be included to reduce the scatter as we have done for the multipoles $\ell$.

\section{Beyond Bispectrum: Shapes of Weak Lensing Trispectrum}
\label{sec:append_tri}
The near all-sky weak lensing maps from surveys such as \textit{Euclid} will also allow
determination of non-Gaussianity statistics beyond the lowest-order, e.g., the
fourth-order correlator or the trispectrum. The trispectrum can be useful not only to
construct the covariance of the power spectrum estimator but also as a consistency check for
the lower order estimators. In this section we will extend the estimator presented
above for the bispectrum to the case of trispectrum.

The trispectrum $T^{\ell_1\ell_2}_{\ell_3\ell_4}(L)$ is defined through the following expression \citep{WH}:
\bes
\ben
&& \la\kappa_{\ell_1 m_1}\kappa_{\ell_2 m}\kappa_{\ell_3 m_3}\kappa_{\ell_4 m_4}\ra  = \sum_{m_i}\sum_{M} (-1)^M
\left ( \begin{array}{ c c c }
     \ell_1 & \ell_2 & L \\
     m_1 & m_2 & M
\end{array} \right)\left ( \begin{array}{ c c c }
     \ell_3 & \ell_4 & L \\
     m_3 & m_4 & -M
\end{array} \right) T^{\ell_1\ell_2}_{\ell_3\ell_4}(L);\\
&& T^{\ell_1\ell_2}_{\ell_3\ell_4}(L) = (2L+1) \sum_{m_i}\sum_{M}\left ( \begin{array}{ c c c }
     \ell_1 & \ell_2 & L \\
     m_1 & m_2 & M
\end{array} \right)\left ( \begin{array}{ c c c }
     \ell_3 & \ell_4 & L \\
     m_3 & m_4 & -M
\end{array} \right) \la\kappa_{\ell_1 m_1}\kappa_{\ell_2 m}\kappa_{\ell_3 m_3}\kappa_{\ell_4 m_4}\ra.
\een
The Wigner 3j-symbols above ensure that the triangle inequalities imposed by statistical isotropy and homogeneity of
the trispectrum in the harmonic space is represented by a quadrilateral.
The indices (harmonics) $\ell_1,\ell_2,\ell_3,\ell_4$ represents the sides of the quadrilateral and  the index $L$
represents one of the diagonal of the quadrilateral.

Using notations introduced before an estimator for the trispectrum can be expressed in the following form:
\ben
&& {\hat T}^{\ell_1\ell_2}_{\ell_3\ell_4}(L) = {2L+1 \over 4\pi} {1 \over h_{\ell_1\ell_2L}}{1 \over h_{\ell_3\ell_4 L}} \sum^{N_{\rm pix}}_{p_1}\sum^{N_{\rm pix}}_{p_2}
{\cal L}_1(\Omega_{p_1}){\cal L}_2(\Omega_{p_1}) {\cal L}_3(\Omega_{p_1}){\cal L}_4(\Omega_{p_2}) P_L(\Omega_{p_1}\cdot\Omega_{p_2}).
\label{eq:tri}
\een
\ees
Here $P_L$ represents Legendre polynomial of order $L$ and $\oh_{p_1}\cdot\oh_{p_2}= \cos\theta_{12}$ where $\theta_{12}$ is the
angle between the two unit vectors $\oh_{p_1}$ and $\oh_{p_2}$ representing the pixels $p_1$ and $p_2$.
The presence of a double sum in Eq.(\ref{eq:tri}) makes it computationally
demanding. However, estimates from degraded low resolution maps can be performed at a reasonable cost. 
The {\em disconnected} part of the trispectrum can be subtracted by using Gaussian maps that are generated
using the same power spectrum even in the absence of noise. The contribution from (Gaussian) noise can be subtracted
in an analogous manner.

Among various configurations that are typically pursued in the literature for the trispectrum
are one-leg squeezed trispectrum, two-leg squeezed trispectrum and collapsed and flattened trispectrum.
The positive trispectrum represents positive kurtosis and negative trispectra represent a negative kurtosis.
Gravitational collapse generates a positive trispectrum.

Few comments are in order. The {\em trapezium} configuration represented by $\ell_1=\ell_2=\ell_3=\ell_4/2$ and the {\em kite} configurations;
represented by $2\ell_1=2\ell_2=\ell_3=\ell_4$ generalise the bispectrum configurations for a positive amplitude they
correspond to elongated over-dense collapsed filament like structure surrounded by underdense regions.
For the squeezed configurations the {\em diagonally squeezed} configurations correspond to $L < \ell_1, \ell_2, \ell_3, \ell_4$
and correspond to modulation of small scale power spectrum by large scale fluctuations.
The {\em one legged squeezed} configuration represents large scale modulation of local or small-scale bispectrum and the two-legged
squeezed configuration correspond to the case when bispectrum is zero and trispectrum is the leading order non-Gaussianity.
In addition flattened configurations of trispectrum which corresponds to $\ell_1=\ell_2=\ell_3 = \ell_4/3$
provides signature that correspond to symmetric perturbations along a line in the real space very similar
to a cosmic string. These configurations of the lensing induced CMB trispectrum has been studied
in some detail \citep{Lewis}. The corresponding investigation in the context of weak lensing at low redshift requires a
numerical fitting function for the trispectrum beyond what can be achieved by perturbative approach.
However, at present there is no suitable fitting formula to derive analytical expressions beyond the perturbative regime
or its extensions using effective field theory (EFT) so we lack the
corresponding formula for the weak lensing trispectrum.
\end{document}